\documentclass[aps,prb,twocolumn,nofootinbib,showpacs,preprintnumbers,notitlepage,groupedaddress]{revtex4-2}

\usepackage{graphicx}
\usepackage{graphics}
\usepackage{amsmath}
\usepackage{amssymb}
\usepackage{amsfonts}
\usepackage{dsfont}
\usepackage{braket}
\usepackage{color}
\usepackage{braket,slashed}
\usepackage[mathscr]{euscript}
\definecolor{darkblue}{rgb}{0, 0, 0.8}
\usepackage[colorlinks=true, breaklinks=true, linkcolor=red, citecolor=blue, urlcolor=blue]{hyperref} 
\usepackage{hyperref}
\usepackage{subfigure}
\usepackage{xfrac}
\usepackage{bm}
\usepackage{kantlipsum}
\usepackage{enumitem}
\usepackage{tikz}
\usepackage{framed}
\usepackage{graphicx}
\usepackage{subfigure}
\usepackage{cleveref}
\usepackage{todonotes}

\allowdisplaybreaks[1]

\newcommand{\code}[1]{\texttt{#1}}

\DeclareMathOperator{\tr}{tr}

\newcommand{\SU}{\ensuremath{\mathrm{SU}}}
\newcommand{\Z}{\ensuremath{\mathbb{Z}}}

\newcommand{\diagram}[2]{\;\vcenter{\hbox{\includegraphics[scale=0.36,page=#2]{./Diagrams/#1.pdf}}}\;}

\begin{document}

\title{The Haldane gap in the SU(3) [3 0 0] Heisenberg  chain}

\author{Lukas Devos}
\email{Lukas.Devos@uGent.be}
\author{Laurens Vanderstraeten}
\author{Frank Verstraete}
\affiliation{Department of Physics and Astronomy, University of Ghent, Krijgslaan 281, 9000 Gent, Belgium}

\begin{abstract}
We calculate the Haldane gap of the $\SU(3)$ spin $[3~0~0]$ Heisenberg model using variational uniform fully symmetric $\SU(3)$ matrix product states, and find that the minimal gap $\Delta /J = 0.0263 $ is obtained in the $[2~1~0]$ sector at momentum $2\pi/3$.  We also discuss the symmetry protected topological order of the ground state, and determine the full dispersion relation of the elementary excitations and the correlation lengths of the system.

\end{abstract}

\maketitle

\par\noindent\emph{\textbf{Introduction---}} %
%
Tensor networks provide variational wavefunctions for approximating the ground states of generic quantum many-body systems in a way that is size-extensive: Local tensors serve as the building blocks for constructing global wavefunctions directly in the thermodynamic limit. This extensivity is also reflected in the symmetry properties, because a global symmetry is encoded as a symmetry of the local tensors that make up the state. This particular feature allows us to classify different phases of matter by characterizing the symmetry properties of these local tensors \emph{and} gives rise to very efficient numerical algorithms by explicitly encoding the symmetry constraints in the tensors and their numerical manipulations \cite{Cirac2021}. 
\par
\begin{figure}[ht!]
\centering
    \includegraphics[width=0.8\columnwidth]{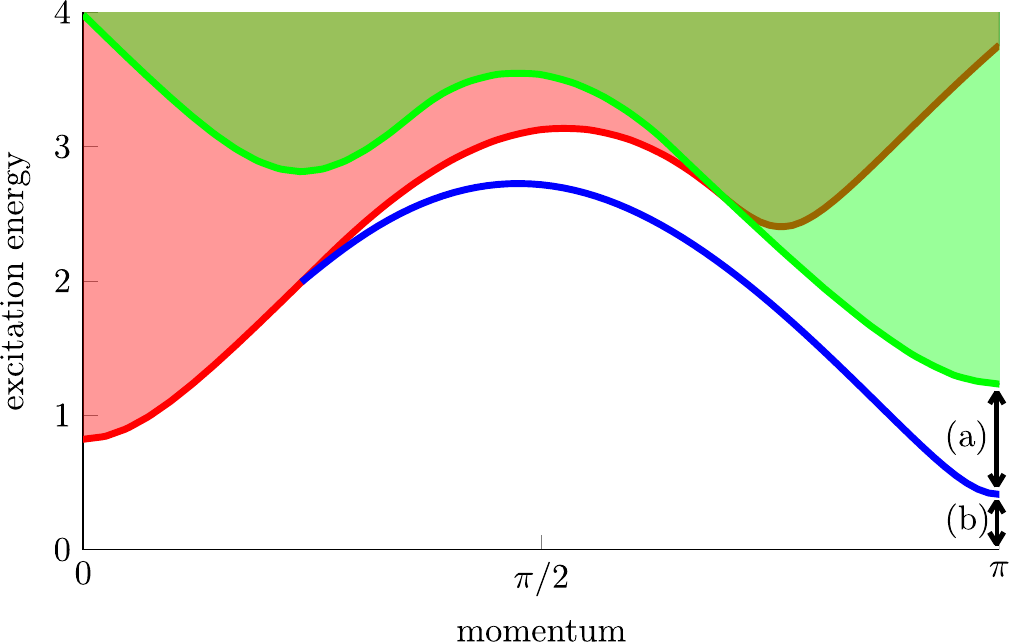}
    \caption{Dispersion relation for the 1-, 2-, and 3-magnon excited states in the $\SU(2)$ spin 1 Heisenberg model, respectively in blue, red and green, reproduced from Ref.~\onlinecite{Vanderstraeten2017}. The excited states are captured with local operators that have an extent that becomes exponentially small with increasing minimum separation between the modes. Around the minimum, this is indicated by the minimum of (a) and (b) and shows that the quasiparticle ansatz is essentially exact \cite{Haegeman2013b}.}
    \label{fig:modes}
\end{figure}

 In one dimension, matrix product states (MPS) provide a framework for simulating the ground-state properties of quantum spin chains and electrons directly in the thermodynamic limit, and generic (non-abelian) symmetries can be exploited in all numerical algorithms \cite{McCulloch2002}. Moreover, this MPS framework can be extended to also capture the low-energy dynamics around the ground state: Interpreting the class of MPS as a variational manifold embedded in the full many-body Hilbert space, the lowest-lying excitations are parametrized by the \emph{tangent space} on this manifold \cite{Haegeman2013a, Vanderstraeten2019}. Indeed, the tangent vectors around the MPS ground state have a natural interpretation as quasiparticles living on top of the strongly-correlated background \cite{Vanderstraeten2015}. These states can then be used as variational ansatz states for the low-lying quasiparticle excitations \cite{Haegeman2013a}; they live in a specific momentum sector and, in combination with the use of symmetries in the MPS, allows us to target definite quantum numbers for the excited states above the ground state \cite{ZaunerStauber2018}. For isolated modes in the spectrum (for example see Fig.~\ref{fig:modes}), it was proven that this local description of the excited-state wavefunction is essentially correct \cite{Haegeman2013b}. In recent years, this quasiparticle ansatz has been used to simulate the low-energy excitation spectrum of generic 1-D \cite{Bera2017, Vanderstraeten2018} and quasi 1-D \cite{VanDamme2021} systems.
\par In this paper, we apply this formalism to simulate the low-energy features of a particular $\SU(3)$ Heisenberg chain. The numerical study of $\SU(N)$ spin chains has gained a new impetus in the last years due to both theoretical and experimental progress. Forty years ago, Haldane made a prediction on the gapped or gapless nature of spin chains depending on which $\SU(2)$ representation is realized on each site in the lattice \cite{Haldane1983a, Haldane1983b, Affleck1989}. Recently, similar predictions have been made for generic $\SU(N)$ chains \cite{Lajko2017, Wamer2020, Affleck:2021jls}. In addition, the Affleck-Kennedy-Lieb-Tasaki construction \cite{Affleck1987, Affleck1988}, which provided a strong motivation for Haldane's prediction, has been extended to some $\SU(N)$ spin chains \cite{Greiter2007, Rachel2009, Morimoto2014, Wamer2019, Gozel2019}. On the experimental side, cold atomic gasses have been used to create $\SU(N)$-symmetric fermionic systems for values as large as $N=10$ \cite{Gorshkov2010}, realising Luttinger liquids, Mott-insulating phases and symmetry-protected topologically ordered (SPT) phases in the lab \cite{Capponi2016, hofrichter_direct_2015, Capponi2020}.
\par Concretely, we study the $\SU(3)$-invariant spin $[3~0~0]$ chain, as defined by Eq.~\eqref{eq:ham}, which is the ``simplest'' case for which a gapped spectrum has been predicted. In Ref.~\onlinecite{Gozel2020}, finite-size MPS methods and intricate extrapolation techniques were used to show the presence of a small gap in this model in the $[3~0~0]$ sector, implying a minimal gap $\Delta / J \in [0.017,0.046]$, confirming the field-theory prediction. Here we use MPS tangent-space methods to simulate the model directly in the thermodynamic limit. We determine the SPT phase of the model, extract reliable estimates for the correlation length and energy gap, and we compute the full dispersion relation, where we have access to all sectors, in contrast to previous studies.

\par\noindent\emph{\textbf{Ground state---}} %
%
In a tensor-network representation, global symmetries are realised by local symmetric tensors which obey ``pulling-through'' types of equations, where the action of an arbitrary group element $g \in \SU(3)$ on a single leg is realised by a representation $U_g$, and can be pulled through to the action on the other legs via the representation $v_g$, as illustrated by Eq.~\eqref{eq:eq1}. This results in a block structure of the tensors, labeled by the irreducible representations that live on their legs.
\par The Hamiltonian for the Heisenberg model is defined as
\begin{equation}\label{eq:ham}
    \mathcal{H} = 2J\sum_i \mathbf{T}_i\mathbf{T}_{i+1} = 2J \diagram{p1}{10}\,,
\end{equation}
where $\mathbf{T}$ are the generators of $\SU(3)$ in the 10-dimensional representation $[3~0~0]$, and the normalisation is fixed by imposing $\tr(T^a T^b) = \frac{15}{2}\delta^{ab}$. While a single generator $T^a$ is not invariant under arbitrary transformations $g$, the vector of all generators transforms under the adjoint representation. As such we construct $\mathbf{T}$ as a three-legged tensor, where the physical legs carry the representation $[3~0~0]$, while the auxiliary leg carries $[2~1~0]$. By virtue of the Mermin-Wagner theorem \cite{Mermin1966}, the ground state leaves the global continuous symmetry unbroken, and combined with the fundamental theorem of MPS \cite{Cirac2021, perez-garcia2007matrix} this leads to a symmetry constraint on the local MPS tensors of the form
\begin{equation}\label{eq:eq1}
    \diagram{p1}{8} = \diagram{p1}{9}\,.
\end{equation}
where now $v_g$ represents a (possibly projective) representation of $g \in \SU(3)$. 
\par The resulting states can be partitioned into three distinct SPT phases, classified by an element of the second cohomology group of $\SU(3)$, which is isomorphic to $\Z_3$ \cite{Morimoto2014, Duivenvoorden2013}. These phases can equivalently be characterised by edge modes that transform under a (combination of) irreducible representations $[m~n~0]$ with fixed value of $m + n \mod 3$, which is reflected in the local symmetry of the MPS tensors by the same restriction on virtual representations $v_g$. Again, the use of symmetric tensors allows us to impose which sectors are present on the virtual level, targeting states that belong to a definite SPT phase. 
\par In order to determine the SPT phase of the ground state, we separately optimise states in each phase, and find that the two non-trivial phases yield equivalent results. Using uniform MPSs directly in the thermodynamic limit, we optimise with a combination of the VUMPS algorithm \cite{ZaunerStauber2018} and Riemannian optimization over the Grassmann manifold of left-canonical MPS tensors, as described in Ref.~\onlinecite{Hauru2021}. We additionally allow for a dynamic distribution of the total effective bond dimension $D$ over the different representations on the virtual level up to a fixed truncation error of the resulting MPS.
\par
\begin{figure}[ht!]
    \includegraphics[width=\columnwidth]{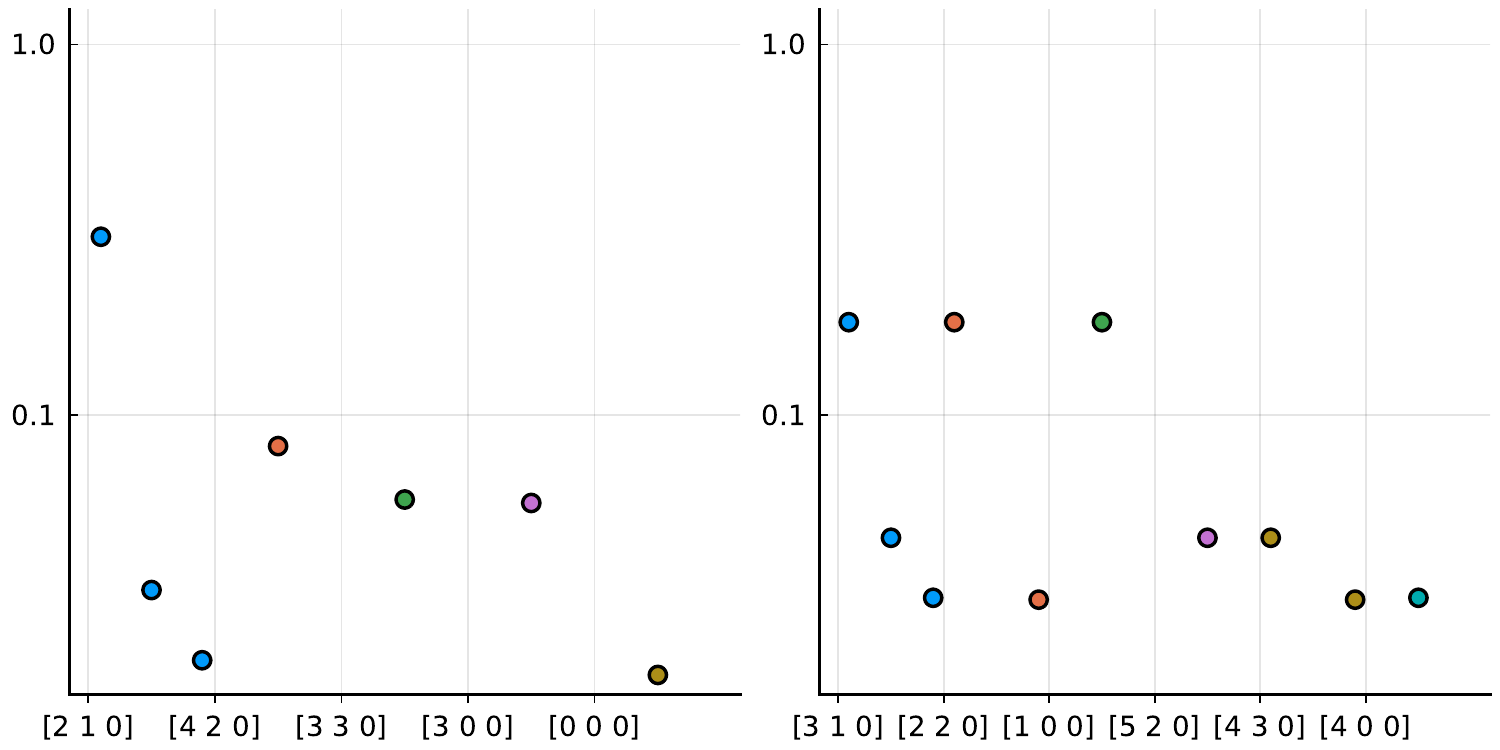}
    \includegraphics[width=0.5\columnwidth]{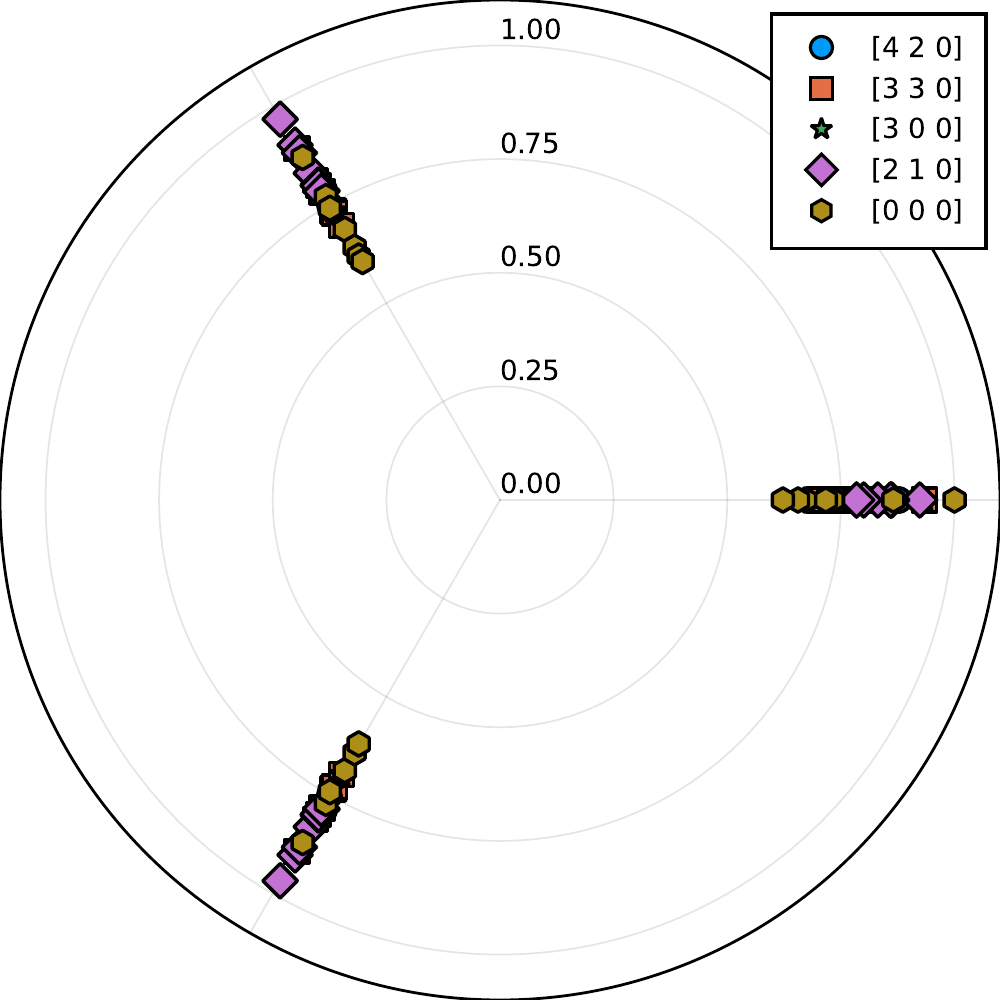}
    \caption{(top) Highest values in the entanglement spectra of the lowest energy state in the trivial SPT phase (left) and the non-trivial SPT phase (right). (bottom) Transfer matrix spectrum of the trivial SPT state.}
    \label{fig:spt}
\end{figure}

 First, we note that for a fixed truncation error the states within the non-trivial SPT phases comparatively take on effective values of $D$ that are about three times larger, while yielding higher values for the energy, showing that they are less suited as a variational ansatz. 
Second, inspection of the entanglement spectrum shows a consistent three-fold degeneracy of all Schmidt values for the non-trivial SPT states, as shown in Fig.~\ref{fig:spt}, which may be interpreted as a consistent matching of charges to imitate an MPS within the trivial SPT phase. For example, singling out the dominant values of both spectra, the decomposition of the tensor product of representations shows $[2~1~0] \in [3~1~0] \otimes [2~2~0] \otimes [1~0~0]$.
Finally, the non-trivial SPT states consistently converge to a non-injective MPS, as is indicated by the degeneracy of the dominant eigenvalue of the transfer matrix, showing that these MPSs are unphysical.
\par To obtain an accurate value of the ground state energy, we compute the lowest energy state for values of the truncation error up to $10^{-6}$. This leads to a variational upper bound on the ground state energy of $E_0 / J = -2.1763966502$ at $D = 122231$, in excellent agreement with the results of Ref.~\onlinecite{Gozel2020}. Furthermore, we linearly extrapolate our results as a function of the variance of the computed states, which yields $E_0 / J = -2.1763984(9)$.
\par We also note that both the entanglement spectrum, as well as the transfer matrix spectrum have degenerate values in conjugate representations.
\newpage
\par\noindent\emph{\textbf{Correlation length---}} %
%
Spin chains whose energy spectrum has a non-vanishing gap above the ground state are characterised by exponentially decaying correlation functions having a finite correlation length $\xi$, while gapless models give rise to power-law behaviour, with infinite correlation length. As such, a finite value of $\xi$ is a strong indicator of a finite energy gap. The correlation length of an MPS with bond dimension $D$ can be calculated as
\begin{equation}\label{eq:xi}
    \xi_D = 1/\epsilon,\qquad\text{ where }\epsilon = -\log |\lambda_1 |
\end{equation}
is the magnitude of the second largest eigenvalue of the MPS transfer matrix (normalised to have largest eigenvalue 1). Additionally, the corresponding complex phase determines the oscillatory behaviour of the correlation function. We note that this will always lead to a finite result at finite $D$, thus we need to extrapolate and show that $\xi_\infty$ remains finite. Adapting the extrapolation procedure outlined in Ref.~\onlinecite{Rams2018}, $\epsilon$ scales with the logarithm of the ratio of the second and third largest eigenvalues of the transfer matrix spectrum as 
\begin{equation} \label{eq:epsdelta}
    \epsilon = a \delta^b + \epsilon_\infty,\qquad\text{ where }\delta=\log\frac{| \lambda_1 |}{| \lambda_2 |}.
\end{equation}
For our purposes, it is sufficient to set $b = 1$, as a linear fit yields excellent results.
\par Imposing the symmetry of the local tensors again allows us to target the eigenvalues of the transfer matrix for eigenvectors with definite quantum numbers, corresponding to elementary excitations carrying those charges. The distribution of the largest magnitude eigenvalues can be seen to have distinct phases, appearing in multiples of $2\pi/3$, and the spectrum is degenerate for conjugate charges. (see Fig.~\ref{fig:spt}).
\par
\begin{figure}
\includegraphics[width=0.9\columnwidth]{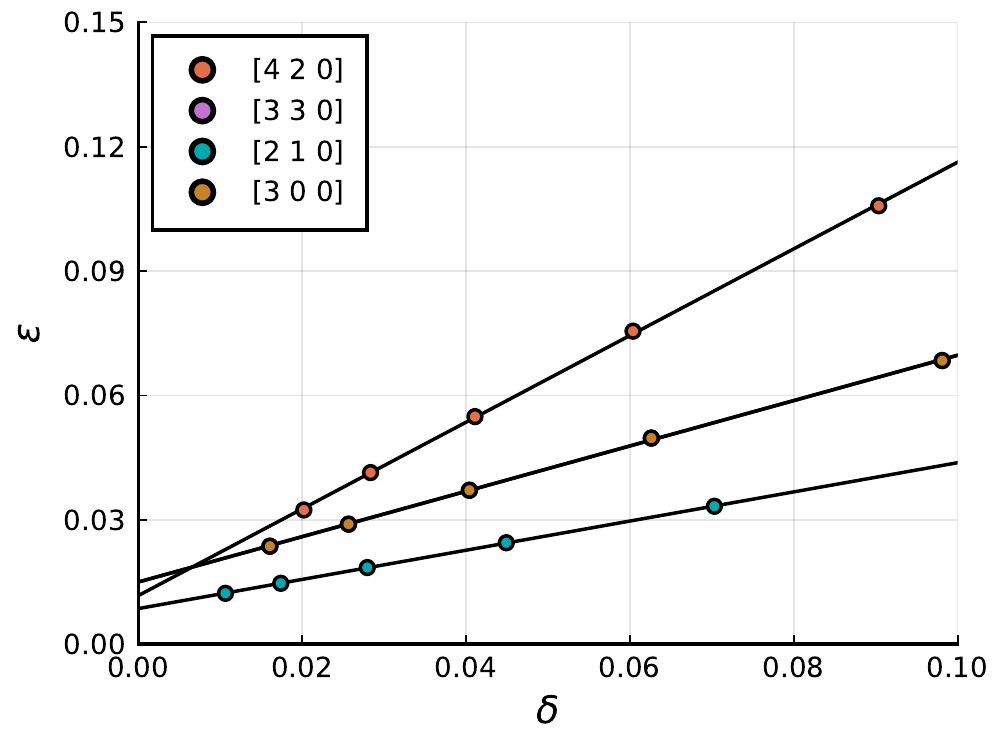}
\caption{\label{fig:epsdelta}Correlation length extrapolations for the ground state. The inverse correlation length $\epsilon$ is plotted against the refinement parameter $\delta$, extrapolated to $\delta = 0$. The conjugate representations give equal results because of the degeneracy of the transfer matrix spectrum.}
\end{figure}

 The resulting extrapolations, which are depicted in Fig.~\ref{fig:epsdelta}, lead to the correlation lengths shown in Tab.~\ref{tab:correlation}, where we note that the dominant correlation length is $\xi = 116.5(8)$ sites for correlation function operators with charge $[2~1~0]$, with an oscillation period of $\pm2\pi/3$.
\begin{table}
\begin{ruledtabular}
\begin{tabular}{ccc}
 charge & correlation length & oscillation period \\
\colrule
$[3~0~0]$ & $66(1)$ & $0$ \\
$[3~3~0]$ & $66(1)$ & $0$  \\
$[2~1~0]$ & $116.5(8)$ & $\pm2\pi/3$\\
$[4~2~0]$ & $85(3)$ & $\pm2\pi/3$
\end{tabular}
\end{ruledtabular}
\caption{Extrapolated correlation length and correlation function oscillation periods.} \label{tab:correlation}
\end{table}

\par\noindent\emph{\textbf{Excitations---}} %
%
\begin{figure}
\includegraphics[width=\columnwidth]{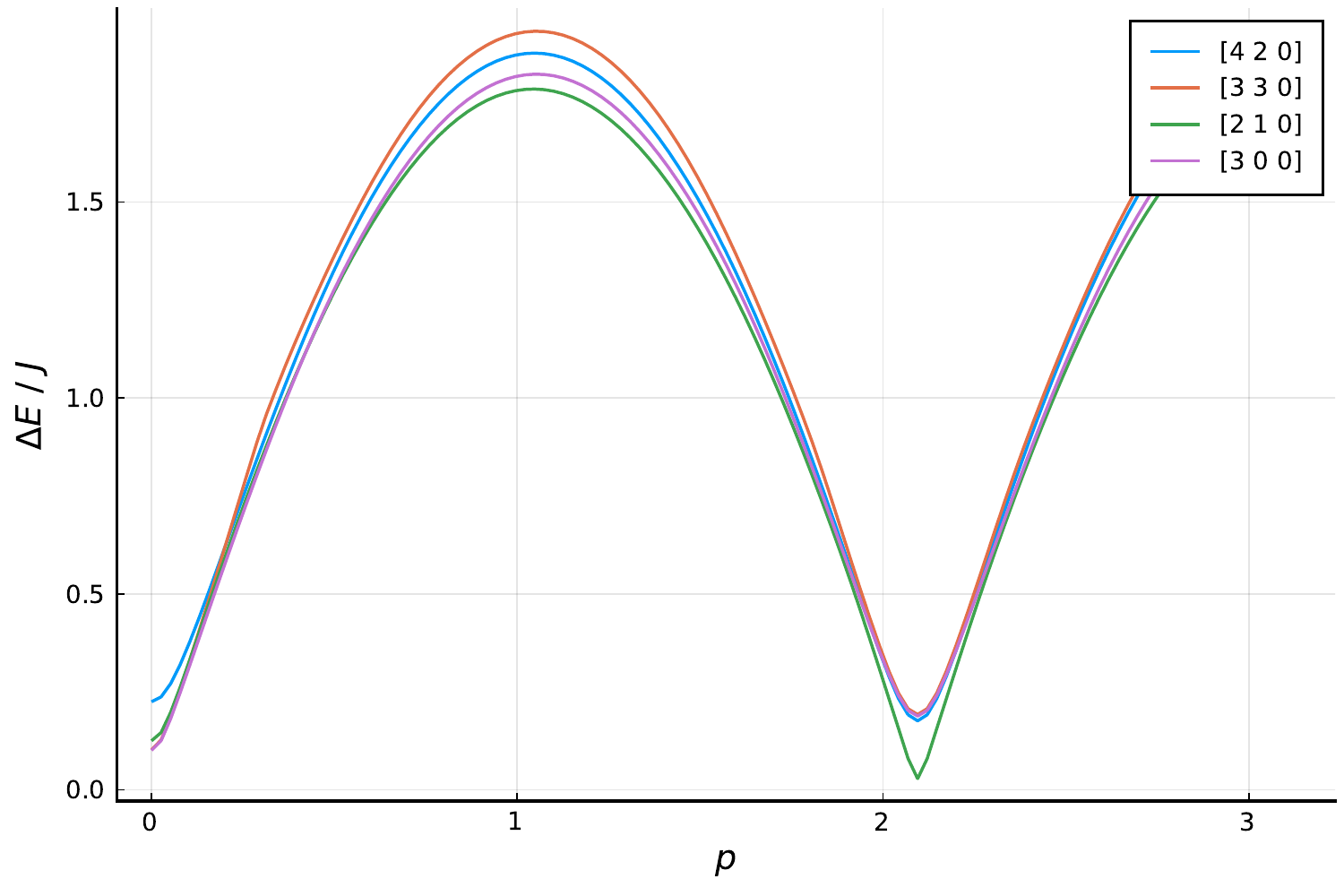}
\includegraphics[width=0.49\columnwidth]{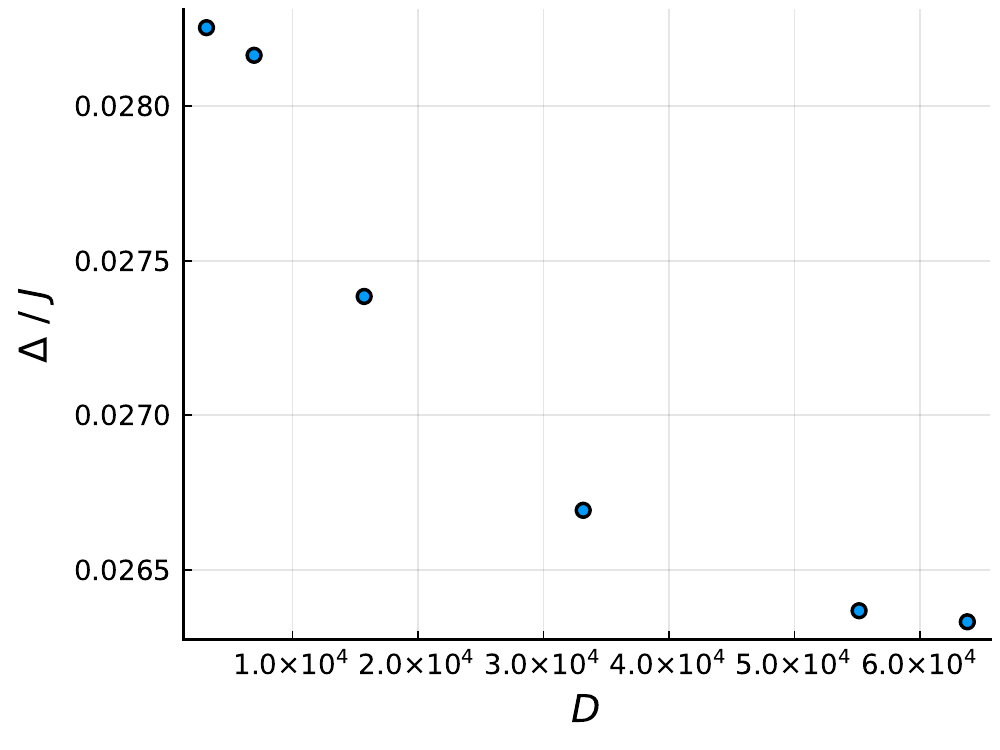}
\includegraphics[width=0.49\columnwidth]{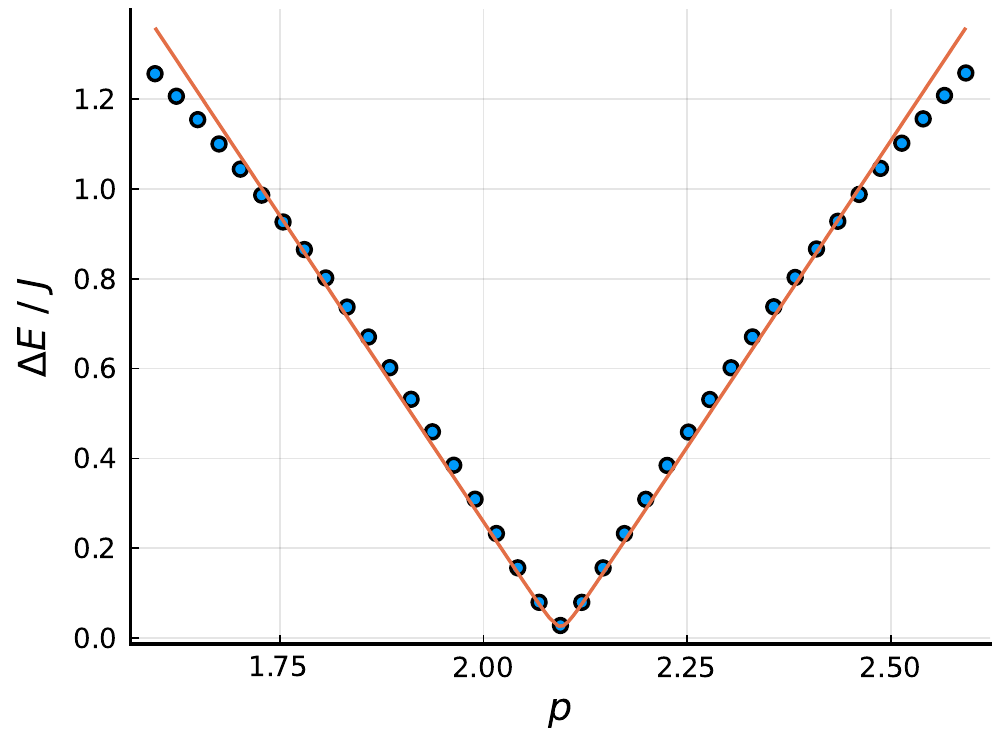}
\caption{(top) Dispersion relation of the elementary excitations of different charges. (bottom-left) Value of the smallest Haldane gap, in the sector $[2~1~0]$, for different bond dimensions $D$. (bottom-right) Calculated dispersion relation around the minimum, along with the theoretical prediction for Lorentz-invariant gapped systems from Eq.~\eqref{eq:dispersion}.\label{fig:disp}}
\end{figure}
As outlined in the introduction, the low-lying elementary excitations can be characterised in terms of the tangent space of the ground state manifold, which is translated into a variational ansatz commonly referred to as the quasiparticle ansatz \cite{Haegeman2012}. This ansatz represents states with definite momentum, and through the use of symmetries can be used to effectively target isolated bands in the spectrum with definite quantum numbers \cite{ZaunerStauber2018}. The ``Haldane gaps'' are expected to be found in sectors that appear in the tensor product of two adjoint representations \cite{Gozel2020}: 
\begin{equation}
[2~1~0] \otimes [2~1~0] \rightarrow 2\cdot[2~1~0] \oplus [3~0~0] \oplus [3~3~0] \oplus [4~2~0]~.
\end{equation}
With a truncation error of the ground state of $10^{-4}$, we compute the full dispersion relation for the elementary excitations, shown in Fig.~\ref{fig:disp}, which shows that the smallest Haldane gap is situated at momentum $p = \pm2\pi/3$ within the $[2~1~0]$ sector. This is consistent with the aforementioned result of the correlation lengths, where the maximum value is found in that same sector, with the same phase.
\par Further decreasing the truncation error, we are able to obtain results up to an effective bond dimension of $D =63766$, and we compute that the smallest gap of the system is $\Delta/J = 0.0263$.
\par As a consistency check, we verify the prediction that the low-energy behaviour in this system can be described by an effective field theory, yielding a dispersion relation of the form 
\begin{equation}\label{eq:dispersion}
    E(p) = \sqrt{\Delta^2 + \nu^2 (p - p_{\textrm{min}})^2}
\end{equation}
where $\nu$ denotes the characteristic velocity in the system. Exploiting Lorentz invariance, we can relate this velocity to the correlation length and the gap as $\nu=\xi\Delta$  \cite{Sorensen1994, Zauner2018}. This prediction can then be compared to the obtained dispersion relation around the minimum, and supplies proof of the validity of our results, as it ties together the correlation length and excited states, which have been calculated independently (Fig.~\ref{fig:disp}).

\par\noindent\emph{\textbf{Conclusions---}} %
%
We show that using tensor-network methods directly in the thermodynamic limit, where the local tensors obey the symmetry of the system allows us to confirm both the theoretical predictions of a Haldane gap in the $\SU(3)$ spin $[3~0~0]$ chain, as well as the numerical results using finite-size scaling methods presented in Ref.~\onlinecite{Gozel2020}.  We find that the ground state belongs to the trivial SPT phase, with an energy density of $E_0 / J = -2.1763984(9)$. The ground state of this system is shown to have a maximal correlation length of $\xi = 116.5(8)$ in the sector $[2~1~0]$ with a phase of $\pm2\pi/3$. The dispersion relation of the elementary excitations is shown to reach its minimum value of $\Delta / J = 0.0263$ in the same sector, at momentum $p=2\pi/3$.

\par\noindent\emph{\textbf{Acknowledgements---}} %
%
We would like to thank Samuel Gozel and Fr\'ed\'eric Mila for inspiring discussions. This work was supported by the Research Foundation Flanders (FWO) and ERC grant QUTE (647905).

\bibliography{bibliography}

\end{document}